# Superfast and sub-wavelength orbital rotation of plasmonic particles in focused Gaussian beams

Lei-Ming Zhou,[1] Xiaoyu Zhu,[1] Yu Zheng,[2,3] Long Wang,[2,3] Chan Huang,[1] Xiaoyun Jiang,[1] Yuzhi Shi,[4] Fang-Wen Sun,[2,3] Jigang Hu[1*]

[1] Department of Optical Engineering, School of Physics, Hefei University of Technology, Hefei, Anhui 230601, China
[2] CAS Key Lab of Quantum Information, University of Science and Technology of China, Hefei 230026, China
[3] CAS Center for Excellence in Quantum Information and Quantum Physics, University of Science and Technology of China, Hefei 230026, China
[4] Institute of Precision Optical Engineering, School of Physics Science and Engineering, Tongji University, Shanghai 200092, China.
* Author to whom correspondence should be addressed: hujigang@hfut.edu.cn

____________________________________________________________

The use of nanophotonics for optical manipulation has continuously attracted interest in both fundamental research and practical applications, due to its significantly enhanced capabilities at the nanoscale. In this work, we showed that plasmonic particles can be trapped at off-axis location in Gaussian beams assisted by surface plasmon resonance. The off-axis displacement can be tuned at the sub-wavelength scale by the incident light beams. Based on these, we propose that a superfast orbital rotation of particles in continuous-wave laser beam can be realized in tightly focused circularly polarized Gaussian beams. The rotation has a tunable orbital radius at the sub-wavelength scale and a superfast rotation speed (more than $10^4$ r/s in water under common laboratory conditions). Our work will aid in the development of optically driven nanomachines, and find applications in micro/nano-rheology, micro-fluid mechanics, and biological research at the nanoscale.

____________________________________________________________

Optical manipulation uses the mechanical effect of light to manipulate objects, with its most famous application known as optical tweezers[1-4]. It has found applications in many fields such as biology, atomic physics and colloid science[2-4]. With the development of both experiment techniques and optical theories (such as nanophotonics), optical manipulation continuously arouses research interests with pulling force[5,6], lateral force[7-9] and negative/inverse torque[10,11] to provide a diversity of controlled motion of particles[12-15]. These controlled motion at the micro/nano-scale promises applications in micromachines[15-18], high-precision measurement[19-21], advanced nanofabrication[22,23] and optofluidic trapping/sorting[24-26].

Usually, the object/particle is trapped at the center of the light spot in optical trapping. For example, optical tweezers capture and manipulate the particles through a focused laser beam as shown in Figs. 1a and 1b. The motion of translation, oscillation, torsion and spinning can be realized with a high-precision control or a fast speed[19]. However, the orbital motion of the particles is not straight-forward as the off-axis trapping of particle is needed first. It's reported that particles with nonlinearity can be trapped off-axis[27]. Using a femto-second laser Gaussian beam with a strong power/intensity, the off-axis trapped nonlinear particle has been demonstrated and showed fast rotation[28]. The rotation angular momentum of the particle is transferred from the orbital angular momentum of the focused beam. Off-axis tapping and rotation of particles without nonlinearity in continuous Gaussian beam is rarely explored.

In this work, we propose a scheme to generate both trapping and propulsion forces along the radial direction in the common Gaussian beam. Through balancing the two forces, we can trap the particle at the off-axis location and tune the off-axis displacement continuously (as shown in Figs. 1c and 1d), especially in the sub-wavelength scale. Changing the trapping beam from linear polarized beam to strongly-focused circularly polarized beam, we can build an orbital rotation system beyond the diffraction-limit. By comparing with the scheme using particle nonlinearity and femto-second laser beam[27,28], only plasmonic particles and continuous-wave laser beam are used; By comparing with the scheme of using structured beams such as Laguerre-Gaussian beams[29,30] and cylindrical beams[31,32], the off-axis trapping distance and the orbital radius is not limited by the diffraction limit and only Gaussian beams are needed here. Also, our scheme does not need auxiliary orbit fabricated by nanofabrication and the orbital radius can be tuned conveniently and continuously[33,34].

Usually, the particle is trapped in the center of the light spot because of the gradient force points to the center, as shown in Fig. 1b. To trap the particle at an off-axis location in the Gaussian beam as shown in Fig. 1c, a propulsive radial-direction force is needed. This can be realized by the excitation of local surface plasmon (LSP) of metallic particles. We will first formulate the principles here. For Rayleigh particle in the electromagnetic field, the gradient force exerted on the particle[35]

$$\mathbf{F} = \frac{\text{Re}(\alpha)}{4}\nabla|\mathbf{E}_i|^2, \quad (1)$$

where $\alpha = \alpha_0/(1 - \frac{ik^3\alpha_0}{6\pi\varepsilon_m})$ is the effective polarizability of the particle, and $\alpha_0 = 4\pi\varepsilon_m a^3 \frac{\varepsilon_p - \varepsilon_m}{\varepsilon_p + 2\varepsilon_m}$ is the static polarizability, $\varepsilon_m$ is the permittivity of surrounding medium. It is noticed that the direction of gradient force depends on the sign of Re($\alpha$), which can be positive and negative. When Re($\alpha$) takes negative value, we obtain propulsion force (i.e., negative gradient force). For clear understanding the properties of $\alpha$, we approximate $\alpha$ with $\alpha_0$ first (i.e., the radiation correction is omitted). It can be seen that when the particle permittivity $\varepsilon_p$ satisfies $\varepsilon_p + 2\varepsilon_m > 0$, then $\alpha_0 < 0$ and the gradient force is propulsion force in Gaussian light spot (denoted as the green line in Fig. 1d); on the other hand, when $\varepsilon_p + 2\varepsilon_m < 0$, then $\alpha_0 > 0$ and the gradient force is trapping force (denoted as the red line in Fig. 1d). The balance point occurs when two Gaussian beams are irradiated and the gradient trapping and propulsion forces are equal, as shown in Fig. 1d.

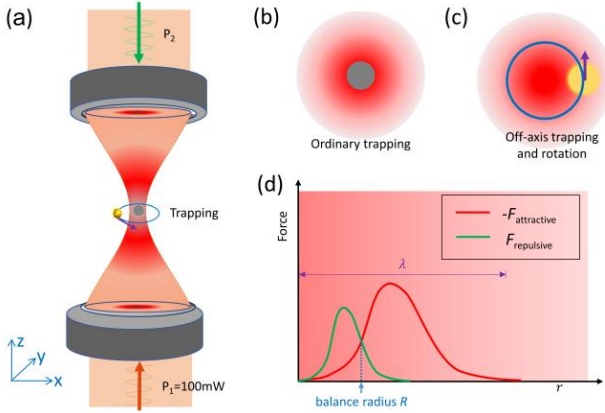

**Fig. 1.** Trapping scheme. (a) Illustration of optical trapping by tightly focused beams. (b) Conventional optical tweezers trap the particles on the axis. (c) Our scheme can trap and rotate metallic particles at off-axis location. (d) The green- and red-light beam can generate propulsive/positive and attractive/negative force in the radial direction, respectively, with the particle trapped at an off-axis equilibrium point in the sub-diffraction scale.

In this work, we will calculate the force and rotation of metallic particles in water. Taking the gold material as an example, the complex polarizability of spherical gold nanoparticle reads

$$\alpha = \alpha' + i\alpha'', \quad (2)$$

where the real and imaginary parts of $\alpha$ are denoted as $\alpha'$ and $\alpha''$, respectively. The polarizability in water has been calculated and plotted in Figs. 2a and 2b, with dependence of vacuum wavelength $\lambda_0$ and particle radius $a$. Here, the dependence of permittivity on wavelength and the expression of $\alpha$ follow Ref. [36] and Ref. [37]. It can be seen that while $\alpha''$ is always positive, $\alpha'$ has negative values. For gold particles with parameters fall into the negative polarizability domain, the gradient force is propulsion force. In other domain, the gradient force is trapping force. The transition between positive and negative domains has been denoted with dash lines in Fig 2a. The transition line is related to the condition of exciting the dipole LSP of plasmonic particles. The excitation condition of LSP of a spherical plasmonic particle is

$$\frac{\varepsilon_p(\lambda_0)}{\varepsilon_m} + \frac{l+1}{l} = 0, \quad (3)$$

where $l$ is the resonant order of LSP. For $l = 1$, we obtain the excitation condition of lowest order LSP (i.e., dipole resonance) $\varepsilon_p + 2\varepsilon_m = 0$, which is just the transition condition of $\alpha_0$. In Figs. 2c and 2d, we show the real and imaginary part of the polarizability of a gold particle with different radii. Taking particle with $a = 40$ nm as an example, it can be seen that at wavelength of 532 nm, $\alpha' < 0$; at wavelength of 1064 nm, $\alpha' > 0$. We choose these two wavelengths to calculate the propulsion and trapping below, because they are also two common wavelengths of lasers.

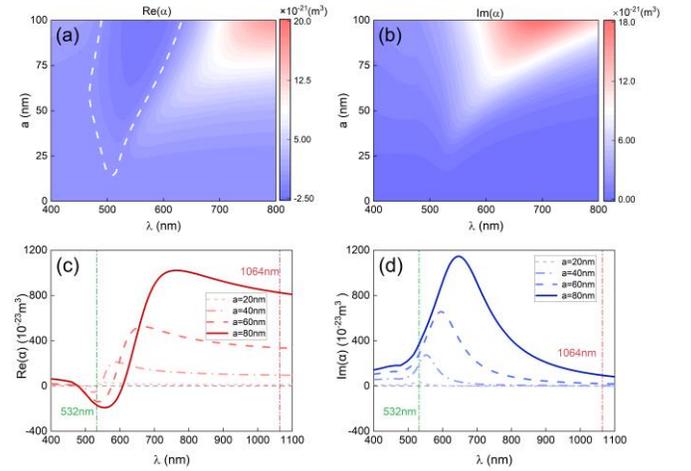

**Fig. 2.** Polarizability of gold particles. (a) The real and (b) imaginary parts of complex polarizability of gold particles with the different wavelengths and particle radii. The white dotted line in (a) denotes the value zero. (c) The real and (d) imaginary parts of the polarizability at several typical radii of $a = 20, 40, 60, 80$ nm.

To show the off-axis trapping, we first investigate the case of two paraxial Gaussian beams. In this case, the trapping field has explicit expression and the polarization of metallic particle takes the Rayleigh approximation. Then the expression for the radial and z-components of the gradient and scattering forces all has approximate explicit expression[37]. We choose two parallel uncoherent counter-propagating beams with wavelengths of 1064 nm (namely Beam 1, with power $P_1$ along the z direction) and 532 nm (namely Beam 2, with power $P_2$ and the opposite z direction), respectively. Both of them have the parameter $s = 0.2$, where $s = \frac{\lambda\sqrt{\varepsilon_m}}{2\pi w_0}$ describes the validity of paraxial beams[37] and $w_0$ is the waist size of the beam (about 1.13 μm and 0.564 μm for the two beams here). As shown in Fig. 3a, the trapping force and the propulsive force balance at $R \approx 280$ nm for a gold particle with $a = 40$ nm when $P_1 = 100$ mW and $P_2 = 20$ mW. If we change the power $P_2$ of the second laser beam, the off-axis trapping location $R$ can be tuned continuously in the sub-wavelength scale. The tunability is shown in Fig. 3b. It's noted that the off-axis displacement is fixed as zero until the power $P_2$ exceeds a certain power (denoted as the transition power $P_t$). This is because both $F_1$ and $F_2$ are promotional to the gradient of the field intensity and hold similar line-shapes. Above the transition power, there are three equilibrium points. The central point is metastable and cannot trap particles under perturbations. Thus, the particle is always off-axis trapped.

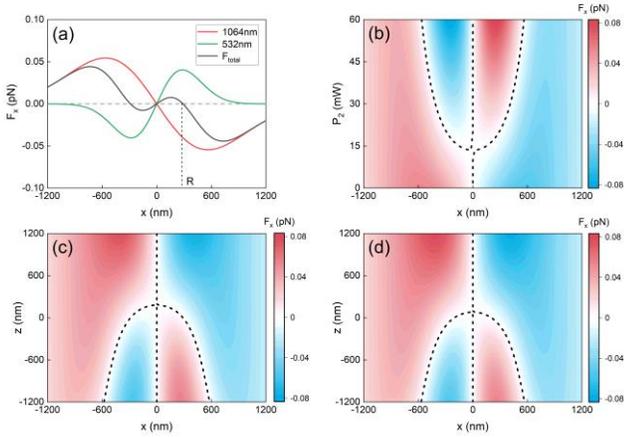

**Fig. 3.** The distribution of radial direction force $F_x$ in two paraxial Gaussian beams of 1064 nm and 532 nm. (a) Force $F_x$ on the focal plane of the 1064 nm and 532 nm laser beam has opposite sign and balance at $R \approx 280$ nm when $P_1 = 100$ mW and $P_2 = 20$ mW. (b) Force $F_x$ with different $P_2$ when $P_1 = 100$ mW. (c) Force $F_x$ on the $xz$ plane when $P_1 = 100$ mW and $P_2 = 20$ mW. (d) The same as Fig. (c), but the waist location of the 532 nm laser beam was shifted to $z_{w2} = 100$ nm.

We also calculate the distribution of radial direction force on the $xz$-plane as shown in Fig. 3c, when $P_1 = 100$ mW and $P_2 = 20$ mW. It can be seen that the trapping radius $R$ decrease with the increment of $z$ and finally goes to zero. This is caused by the convergent or divergent energy flux along radial direction out of the focus plane, which generates an additional scattering force along the radial direction. The participation of scattering force also provides a way to tune the location $R$. As an example, the distribution of total transverse force $F_x$ is shown Fig. 3d, when the waist location of Beam 2 is moved from $z_{w2} = 0$ to $z_{w2} = 100$ nm.

Below we show our accurate numerical simulation result of off-axis trapping by strongly focused beams and the ultrafast orbital rotation of the gold particles. In the experiment of optical manipulation, strongly focused beams are used usually where the high numerical aperture lenses are introduced. The beam is no more paraxial, and field distribution near the focus is distorted[38]. Specifically, the spin angular momentum (SAM) can be conversed to orbital angular momentum (OAM) with focusing, which can further drive the orbital rotation of off-axis trapped particles[28]. The tightly focused beam was modeled by vector diffraction theory and calculated with Debye-Wolf integral[39]. The forces exerted on particles were then calculated with T-matrix method[39].

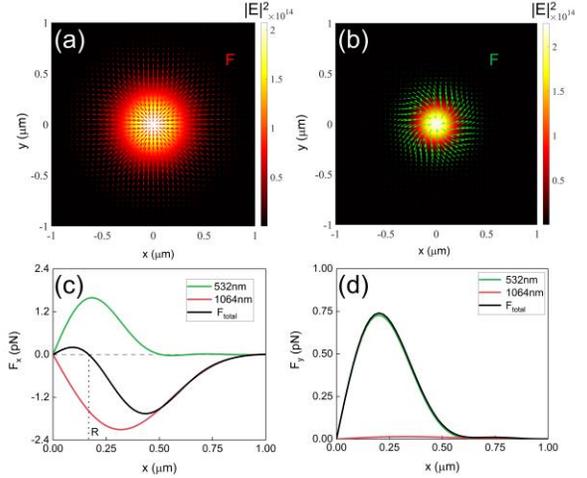

**Fig. 4.** Force field of the gold particle on the focal plane in a tightly focused circularly-polarized Gaussian beam. (a) For laser beam with wavelength 1064 nm and power $P_1 = 100$ mW. (b) For laser beam with wavelength 532 nm and power $P_2 = 40$ mW. The green and red arrows denote the force field $F$. (c) The forces in the radial direction for two beams. (d) The forces in the angular direction for two beams.

We show the simulation result for the force field on the focal plane in circularly-polarized beams as an example, as shown in Fig. 4. For the first beam with a wavelength of 1064nm and left-circularly polarization, the transverse force (denoted by red arrows) is attractive force in the radial direction and shows rotating force component in the clockwise direction (Fig. 4a). For the second counter-propagating beam with wavelength 532 nm and right-circularly polarization, the transverse force (denoted by green arrows) is repulsive force in the radial direction and shows also rotating force component in the clockwise direction (Fig. 4b). The field intensity distribution on the focal plane is also shown in Figs. 4a and 4c. The parameters are: $a = 40$ nm; $P_1 = 100$ mW, $P_2 = 40$ mW; $NA_1 = 1.2, NA_2 = 1.0$. For the total force applied on gold particles when both laser beams are power-on, the $x$ component (trapping force) and $y$ component (rotational force) are shown in Figs. 4c and 4d, respectively. At the location $R$ in the radial direction, the total $F_x$ is zero and the particle is trapped; at the same time, the total $F_y$ induced by OAM conversed from SAM of light is non-zero and will rotate the particle in the orbit. Further calculations show that the trapping is stable with thermal perturbations (see Supplementary Material). In the focusing field, the off-axis trapping position (i.e., the radius of the rotation orbit) is at the sub-diffraction scale.

In the last part, we investigate the tunability and figure of merit of this orbital rotation system. The orbital radius $R$ of the gold particles can be tuned by the power $P_2$ or the numerical aperture $NA_2$ of the 532 nm laser. For the tuning by power $P_2$ (Fig. 5a), the transition power $P_t$ is larger than that in the paraxial case, and the radius $R$ also becomes smaller. The transverse force $F_x$ in the $xz$ plane is shown in Fig. 5b when $P_1 = 100$ mW and $P_2 = 40$ mW. The orbital radius $R$ is about 150 nm in the focal plane. We then investigated R in the focal plane by changing $NA_2$ in Fig. 5c while other parameters are the same as Fig. 5b. With increasement of $NA_2$, $R$ changes little. However, with a high NA, the OAM conversed from SAM increases and particle can rotate faster because of the increased $F_y$. In Fig. 5d, we simulate the rotation of gold particles in water at the temperature $T = 373.15$ K. The particle rotation velocity is calculated when the rotating force is equal to the viscous force caused by the motion of the particle in water[28]. The rotate speed can reach larger than $5.0 \times 10^4$ r/s if we increase the incident power by ten times.

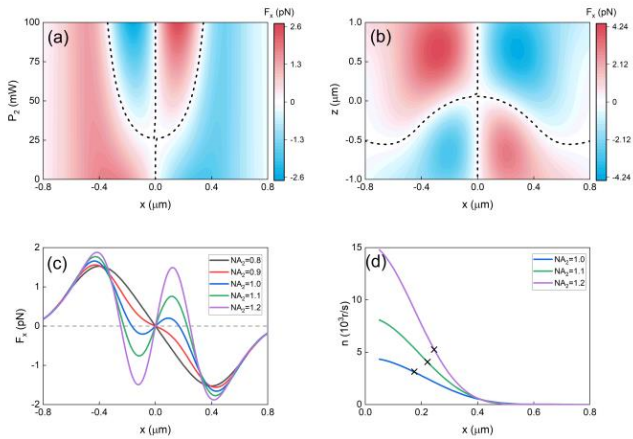

**Fig. 5.** Tunability and rotation speed of this orbital rotation system. (a) The total transverse force $F_x$ with different $P_2$ when $P_1 = 100$ mW. $NA_1 = 1.2$, $NA_2 = 1.0$, and the black dashed curve denotes the equilibrium position. (b) Same as (a), for $F_x$ on the $xz$ plane when $P_2 = 40$ mW. (c) Same as (b), except the changing of $NA_2$. (d) Rotation speeds of gold particles in water correspond to different $NA_2$ in (c). Black × denotes the trapping radius.

As a discussion, it is noted that we discuss only the force in the transverse plane here. The effect of the longitude force (z-direction force) can be balanced through choosing appropriate beam power $P_2$ or just using a substrate to support the particles. Also, the width of the ring-shape orbit is about 100nm when we choose potential depth of $10 \, k_B T$. It is estimated that transverse off-set of z-axes need to be less than several tens of nanometers to keep the orbit. Additionally, the propulsive force in the radial direction caused by divergent energy flux can also be used to trap the particle at the off-axis location even the particle is dielectric as we will discuss it in another independent work. However, the dielectric particles cannot rotate fast because the rotating force is not improved by the LSP resonance of the particle.

In conclusion, this work proposes the off-axis trapping of metallic particles and superfast orbital rotation with ultra-small radii. This proposal does not need complex systems, auxiliary orbits or femto-second laser. The proposed system uses the simple circularly-polarized Gaussian beams, lenses and dielectric substrate, which can help the construction of optically driven nanomachines, and find applications in micro/nano-rheology and micro-fluid mechanics at the nanoscale. This work also shows that the enhanced optical scattering force induced by LSP in metallic particles can be efficiently utilized in advanced optical manipulation.

**Supplementary Material.** See the supplementary material for the simulation and discussion of the potential well depth.

**Acknowledgements.** This work was supported by National Natural Science Foundation of China (12204140, 62075053, U20A20216 and 62205246); Fundamental Research Funds for the Central Universities (JZ2023HGTB0223, PA2020GDKC0024, JZ2022HGTA0340, JZ2022HGQA0201); Natural Science Foundation of Anhui Province (JZ2022AKZR0437); Shanghai Pilot Program for Basic Research, Science and Technology Commission of Shanghai Municipality (22ZR1432400).

**Author Declarations**. The authors have no conflicts to disclose.

**Data Availability.** The data that support the findings of this study are available from the corresponding author upon reasonable request.